# Enhancement of Unruh effect in high-intensity laser fields using hyperbolic metamaterials

**IGOR I. SMOLYANINOV**

*Department of Electrical and Computer Engineering, University of Maryland, College Park, MD 20742, USA*
*smoly@umd.edu*

**Abstract:** High-intensity laser fields acting on free electrons were proposed to create sufficiently large accelerations to enable detection of Unruh radiation. However, the currently achievable electron accelerations are not large enough. Here we demonstrate that Unruh effect in high-intensity laser fields is strongly enhanced near hyperbolic metamaterials, so that many orders of magnitude smaller accelerations may be used to detect Unruh radiation.

Theoretical prediction that an accelerating object perceives its surroundings as a bath of thermal radiation, even if it accelerates in vacuum, is called the Unruh effect [1]. The physical reason for this effect is that the vacuum states for the inertial (Minkowski) observers and the uniformly accelerated (Rindler) observers are defined with respect to a different set of photon wave functions, which belong to different Hilbert spaces [2,3]. As a result, the uniformly accelerated object experiences the fluctuations of the Minkowski quantum vacuum as a thermal bath of particles at the Unruh temperature

$$T_U = \frac{\hbar a}{2\pi k_B c}, \qquad (1)$$

which is defined by its acceleration $a$. Since an observer accelerating at $a=g=9.8$ m/s$^2$ should experience Unruh temperature of only $4\times10^{-20}$ K, this effect is very difficult to detect in the experiment. On the other hand, since Unruh effect plays a very important role in understanding the structure of quantum vacuum in gravitational fields, its experimental detection and study is one of the key challenges of contemporary fundamental physics.

One of the experimental approaches aimed at detection of Unruh radiation uses ultrahigh-intensity laser fields accelerating free electrons [4]. Indeed, due to enormous progress in the development of high-intensity, short-pulse lasers in recent years, the experimental approach to Unruh radiation detection may become quite realistic in the near future. For example, high-power lasers can now reach focused intensities of more than $10^{19}$ W cm-2 at high repetition rates, and they can accelerate electrons to highly relativistic energies of the order of GeV within distances of a few mm [5]. In such an experimental configuration the Unruh radiation power must be compared to the power of Larmor radiation, which is given by the classical expression

$$P_L = \frac{2e^2 a^2}{3c^3} \qquad (2)$$

According to the quantitative analysis in [4], in such experiments the ratio of the powers of Unruh and Larmor radiation reaches $P_U/P_L \sim 10^{-10}$, and further refinement of these experiments may be achieved by taking into account different spatial distributions of Unruh and Larmor radiations (the latter being absent in the acceleration direction), and by taking advantage of the fact that Unruh radiation consists of entangled photon pairs, while Larmor radiation is classical. Regardless of the eventual successful experimental configuration, there is a need to further enhance Unruh effect, so that its experimental detection may become possible.

The power of Unruh radiation $P_U$ in vacuum may be estimated using the following simple quasi-classical arguments [6]. In the instantaneous rest frame of an accelerated electron, $P_U$ may be estimated as a product of energy flux of thermal radiation at the Unruh temperature (defined by Eq.(1)) and the electron-photon scattering cross section. Assuming the Thomson scattering cross section

$$\sigma_{Thomson} = \frac{8\pi}{3} r_0^2 \quad , \tag{3}$$

where $r_0 = e^2/mc^2$ is the classical electron radius, and the energy density of thermal radiation given by the usual Planck expression

$$\frac{dU}{d\omega} = \frac{1}{\pi^2 c^3} \frac{\hbar \omega^3}{e^{\hbar\omega/kT} - 1} \quad , \tag{4}$$

we end up with an estimate of $P_U$ as follows:

$$\frac{dU_{Unruh}}{dt d\omega} = \frac{1}{\pi^2 c^2} \frac{\hbar \omega^3}{e^{\hbar\omega/kT} - 1} \frac{8\pi}{3} r_0^2 \tag{5}$$

and upon integration over $\omega$ we find

$$P_{Unruh} = \frac{8\pi^3 \hbar r_0^2}{45 c^2} \left(\frac{kT}{\hbar}\right)^4 = \frac{\hbar r_0^2 a^4}{90\pi c^6} \tag{6}$$

(using Eq.(1) for $T_U$). Comparison of Eqs. (2) and (6) indicates that $P_U$ would reach the same order of magnitude as $P_L$ in vacuum only at very large accelerations of the order of $10^{30} g$. While it was demonstrated very recently that such giant accelerations may indeed be created in the lab for photon "quasi-particles" [3,7], achieving a similar result with electrons may be more difficult.

Fortunately, we should be able to considerably enhance the energy density of thermal radiation using hyperbolic metamaterials, as illustrated in Fig.1. Indeed, it was demonstrated very recently [8,9] that in the absence of diffraction limit on the photon wave vector in hyperbolic metamaterials, the conventional Planck expression for the thermal energy density (Eq.(4)) is no longer valid, and the thermal energy density inside and near hyperbolic metamaterials increases by many orders of magnitude.

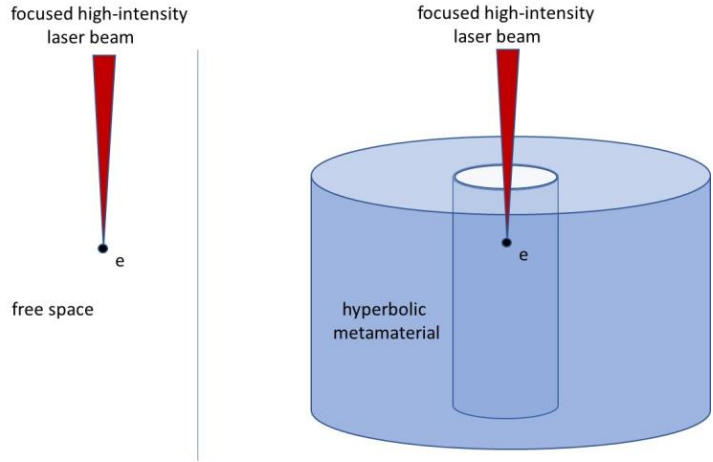

Fig. 1. High intensity laser beam accelerates a free electron in free space (left) and near a hyperbolic metamaterial (right).

Thermal equilibrium in an accelerated reference frame requires a hyperbolic metamaterial to have the same temperature $T_U$ as free space. On the other hand, as discussed in [8,9], since the density of photonic states $\rho(\omega)$ inside a hyperbolic metamaterial is enhanced considerably, the conventional Planck expression for the energy density of thermal radiation (Eq.(4)) needs to be modified. In the most general case, the energy density of thermal radiation must be written as

$$U = \int_0^\infty d\omega \frac{\hbar\omega}{e^{\hbar\omega/kT}-1} \rho(\omega) \qquad (7)$$

In vacuum and in elliptic media $\rho(\omega) \sim \omega^2$, leading to the Stefan-Boltzmann radiative energy flux $\sim T^4$, which leads to Eq.(6). On the other hand, broadband divergence of $\rho(\omega)$ in hyperbolic media may lead to a very large increase in the radiative energy flux [8,9]. Eq.(7) clearly demonstrates that the drastic change in the photon density of states schematically illustrated in Fig.2 must lead to a drastic change in the energy density of thermal radiation.

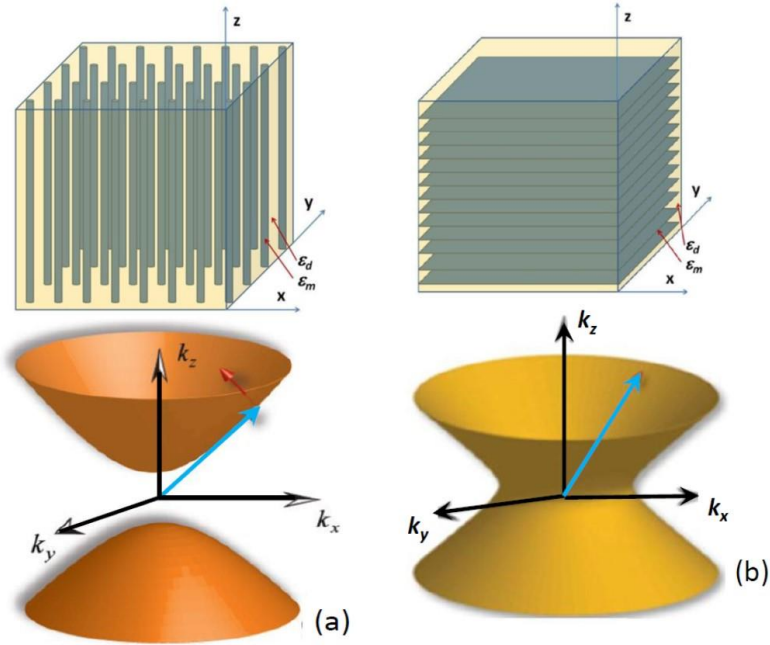

Fig. 2. The constant frequency surfaces in k-space for (a) two-sheet uniaxial hyperbolic metamaterial in which $\varepsilon_z < 0$ and $\varepsilon_{xy} > 0$ (which is typical for metal nanowire-based metamaterials). and (b) one-sheet uniaxial hyperbolic metamaterial in which $\varepsilon_z > 0$ and $\varepsilon_{xy} < 0$ (which is typical for metal nanolayer-based metamaterials). Unlike elliptical media, these surfaces are open. The blue arrow shows a photon wave vector, which is not diffraction-limited, leading to drastic increase in $\rho(\omega)$, which is proportional to the area of the surface of constant frequency.

For the energy flux $S_T$ along the symmetry axis of a uniaxial hyperbolic metamaterial, it was found [8] that

$$S_T = \frac{\hbar c^2 k_{max}^4}{32\pi^2} \int d\omega \frac{1}{\exp\left(\frac{\hbar\omega}{kT}\right)-1} \left| \frac{\varepsilon_1 \frac{d\varepsilon_2}{d\omega} - \varepsilon_2 \frac{d\varepsilon_1}{d\omega}}{\det\|\varepsilon\|} \right|, \qquad (8)$$

where $k_{max} \sim 1/d$ is the wave vector cut-off defined by the metamaterial structural parameter $d$, $\varepsilon_1 = \varepsilon_{xy}$ and $\varepsilon_2 = \varepsilon_z$ are the dielectric tensor components of the hyperbolic metamaterial, and the frequency integration is performed over the frequency bands corresponding to the hyperbolic dispersion. Note that the heat flux in Eq. (8) is very sensitive to the dispersion $d\varepsilon/d\omega$ in the hyperbolic metamaterial. Indeed, the derivative of the dielectric permittivity determines the difference in the asymptotic behaviour of the $k$-vector between the two hyperbolic surfaces that determine the phase space volume between the frequencies $\omega$ and $\omega + d\omega$ (see Fig. 2), and therefore it defines $\rho(\omega)$.

The most practical and widely used hyperbolic metamaterial designs rely on either the metal-dielectric multilayer geometry, or incorporate aligned metal nanowire composites [9]. In both designs the hyperbolic metamaterial behavior is broadband, and it overlaps with the typical thermal wavelength range. Depending on the geometry of the constant frequency surfaces in k-space shown in Fig.2, the resulting energy flux was calculated as follows [8]:

$$S_T \approx \frac{5}{16\pi^2} S_T^{(0)} \left( \frac{k^2_{max}}{k_T k_p} \right)^2 \qquad (9)$$

for the nanowire array metamaterial design, and

$$S_T \approx \frac{\varepsilon_d}{4(1-n)} S_T^{(0)} \left( \frac{k_{max}}{k_p} \right)^4 \qquad (10)$$

for the layered metamaterial design, where $S^{(0)}_T$ is the blackbody thermal energy flux for emission into free space, $k_T = k_B T/\hbar c$ is the typical thermal momentum, $\varepsilon_d$ is the dielectric permittivity of the dielectric component of the metamaterial, and $n$ is the volume fraction of metal nanolayers. In Eqs.(9,10) we also introduced $k_p$ as a typical "plasma momentum" of the metamaterial, which is defined, as usual, as

$$k_p = \sqrt{\frac{4\pi N}{m^*}} \frac{e}{c}, \qquad (11)$$

where $N$ and $m^*$ are the free charge carrier density in the metamaterial and their effective mass, respectively. Numerical analysis demonstrate that in both cases the numerical values of $S_T$ defined by Eqs.(9,10) exceed $S^{(0)}_T$ by many orders of magnitude, since in a typical hyperbolic metamaterial $k_{max} \gg k_p$ and $k_{max} \gg k_T$, which should lead to very large enhancement of Unruh radiation near hyperbolic metamaterials.

Let us start our numerical analysis with the case of a nanowire array metamaterial corresponding to Eq.(9) and Fig. 2(a). Using Eqs.(6,9), we obtain the following power of Unruh radiation near a hyperbolic metamaterial:

$$P^{hyp}_{Unruh} = \frac{\pi r_0^2 k_B^2 T^2}{18\hbar d^4 k_p^2} = \frac{\hbar r_0^2}{72\pi d^4 k_p^2 c^2} a^2 = \frac{r_0^2}{48\pi \alpha d^4 k_p^2} P_{Larmor}, \qquad (12)$$

where $\alpha$ is the fine structure constant. As can be seen from Eq.(12), both $P_U$ and $P_L$ are proportional to $a^2$ in this case. Assuming the most optimistic case of $d \sim 1$ nm and $k_p \sim 1$ μm$^{-1}$ (which are indeed observed in natural hyperbolic metamaterials [10]), the ratio of the powers of Unruh and Larmor radiation reaches $P_U/P_L \sim 10^{-5}$, which is five order of magnitude larger compared to the estimates in free space reported in [4]. In combination with different spatial distributions of Unruh and Larmor radiations, and by taking advantage of the entanglement properties of Unruh photon pairs, this enhancement may lead to a realistic experiment aimed at detection of the Unruh effect. The only potential drawback of this case is that unlike in free space experiments, the $P_U/P_L$ ratio is constant, and one cannot gain from using higher accelerations.

Let us now consider the case of a nanolayer-based hyperbolic metamaterial, which corresponds to Eq.(10) and Fig. 2(b). By comparing Eqs.(6) and (10), it is clear that in this case

$$P_{Unruh}^{hyp} = \frac{\varepsilon_d}{4(1-n)}\left(\frac{k_{max}}{k_p}\right)^4 \frac{8\pi^3 \hbar r_0^2}{45c^2}\left(\frac{kT}{\hbar}\right)^4 = \frac{\varepsilon_d}{4(1-n)}\left(\frac{k_{max}}{k_p}\right)^4 \frac{\hbar r_0^2 a^4}{90\pi c^6}, \tag{13}$$

where the Unruh effect enhancement compared to free space is defined by the $(k_{max}/k_p)^4$ factor. Assuming once again the most optimistic case of $d \sim 1$ nm and $k_p \sim 1$ μm$^{-1}$, this enhancement factor may reach up to $10^{12}$, which means that all other experimental factors being the same, three order of magnitude smaller accelerations will be needed to detect Unruh radiation.

Our numerical results are summarized in Fig. 3, which compares the power of Larmor and Unruh radiation in free space and near hyperbolic metamaterials of the two kinds discussed above.

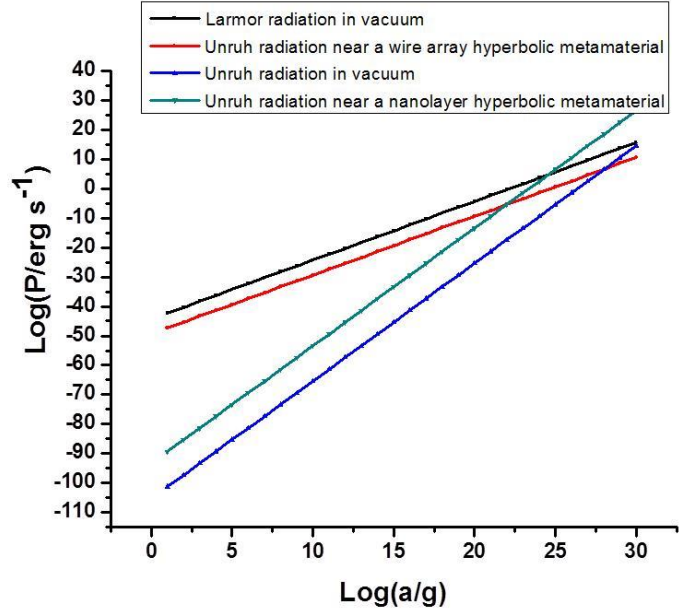

Fig. 3. Comparison of the power of Larmor and Unruh radiation in free space and near hyperbolic metamaterials.

While the Unruh effect enhancement is indeed huge inside and near hyperbolic metamaterials, it is important to understand how close we need to be to a metamaterial, while avoiding direct interaction between an accelerated electron and the solid state constituents of the metamaterial. Since the spectral range of thermal radiation belongs to the long-wavelength IR range, an accelerated electron may propagate in vacuum outside the metamaterial (as for example is shown in Fig.1) while still be located inside the near-field range of thermal radiation of the metamaterial. Moreover, if a nanowire array metamaterial geometry (Fig.2a) is implemented in such an experiment, the free standing nanowires may be used (see for example [11]), so that an accelerated electron may propagate in free space in between the nanowires. In fact, such an experimental geometry will be somewhat similar to the observations of photoluminescence from metal nanotips, which already indicate a potential contribution from the Unruh effect [12].